\documentclass[aps,prl,reprint,superscriptaddress,twocolumn]{revtex4-1}
\usepackage{graphicx}
\usepackage[english]{babel} %
\usepackage[usenames]{color}
\usepackage{subfigure}
\usepackage{dcolumn}
\usepackage{bm}
\usepackage{amsmath,amsfonts,amssymb,float}
\usepackage{natbib}
\usepackage{epsfig}
\usepackage{multirow}
\usepackage{textcomp} 
\usepackage{amsmath} 
\usepackage{amsthm} 
\usepackage{amssymb}	
\usepackage{graphicx} 
\makeatletter 

\let\baraccent=\= 
\renewcommand{\=}[1]{\stackrel{#1}{=}} 

\theoremstyle{definition}

\theoremstyle{remark}

\begin{document}
\title{A focusable, convergent fast-electron beam from ultra-high-intensity laser-solid interactions}
\author{R.H.H. Scott}
\email{Robbie.Scott@stfc.ac.uk}
\thanks{}
\affiliation{Central Laser Facility, STFC Rutherford Appleton Laboratory, Harwell Oxford, Didcot, OX11 0QX, United Kingdom}
\date{\today}

\begin{abstract}
A novel scheme for the creation of a convergent, or focussing, fast-electron beam generated from ultra-high-intensity laser-solid interactions is described. Self-consistent particle-in-cell simulations are used to demonstrate the efficacy of this scheme in two dimensions. It is shown that a beam of fast-electrons of energy 500 keV - 3 MeV propagates within a solid-density plasma, focussing at depth. The depth of focus of the fast-electron beam is controlled via the target dimensions and focussing optics. 
\end{abstract}

\maketitle

The study of fast-electron generation and subsequent transport in high-density plasmas is important for numerous applications including proton and ion beam production \cite{PhysRevLett.84.670}, isochoric heating of high density matter for opacity and equation-of-state studies \cite{Hoarty2007115}, x-ray sources, and fast-ignition inertial fusion \cite{Tabak:1994ov}.

There are a number of schemes to ignite a small mass of inertially-confined Deuterium-Tritium (DT) fuel, initiating a thermonuclear burn wave which propagates through a larger mass of fuel. The most developed inertial confinement fusion (ICF) scheme is central hotspot ignition, in which the spherical implosion of a plastic capsule containing a solid DT layer compresses and heats a small mass of DT gas to $\sim 5$ keV and a density-radius product ($\rho r$) of $0.4$ gcm$^{-2}$. In order to create these hotspot conditions simultaneously via spherical compression, excess energy must be expended compressing the cold DT ice and ablator, limiting the fusion energy gain for more robust central hotspot target designs. The principal alternative laser-driven ICF schemes are shock and fast ignition, these may require less energy to form the hotspot and hence have the potential for higher gain. 

In the fast-ignition scheme, the DT fuel is firstly compressed using long-pulse lasers, then a shorter duration, high-intensity ($\sim 1\times10^{20}$Wcm$^{-2}$) laser is used to generate a beam of fast-electrons, these stop collisionally within the compressed DT ice, heating it. The coupling efficiency from the high-intensity laser to the DT fuel is determined by the fraction of laser energy absorbed into fast-electrons, their energy spectrum, divergence, and the distance from the laser absorption surface to the compressed DT target or core \cite{Atzeni:2005gf}. A number of studies \cite{Green:2008jw} have found the inherent electron beam divergence to be so large that the high-intensity laser energy required for fast-ignition is excessive from both a practical and economic perspective. This in turn has led to proposals \cite{PhysRevLett.108.125004,0741-3335-54-8-085016,Robinson:2008aa,PhysRevLett.109.015001} which seek to address this issue by re-directing the divergent fast-electron beam towards the compressed core, thereby reducing the high-intensity laser energy required for ignition. 

This letter describes a technique to control the divergence of the fast-electron \emph{source}. Particle-in-cell simulations are used to demonstrate that this technique can be used to create a convergent fast-electron source. The fast-electron beam focal length can be controlled simply by altering the focussing optics and interaction geometry, and it is shown that the generated fast-electron beam can be focussed within a solid. 

The Lorentz force equation $\bf{F} = -q(\bf{E} + \bf{v\times B})$ describes the interaction of a single electron with a plane electromagnetic wave, here $q$ is the electron charge, $\bf{E}$ the electric field vector, $\bf{B}$ the magnetic field vector and $\bf{v}$ is the electron's velocity. The electromagnetic wave firstly accelerates the electron in the direction of the electric field, if the magnitude of the electric field is sufficient for the electron to obtain a velocity approaching the speed of light $c$, the $\bf{v}\times \bf{B}$ component of the Lorentz force equation becomes significant with respect to that due to $\bf{E}$, causing the electron to also be accelerated in a direction parallel to the wave's Poynting vector. Defining the normalised vector potential as ${\bf{a_0}} = q {\bf{E}} /m_ec\omega $ where $m_e$ is the electron mass and $\omega$ the wave's angular frequency, we find that for a wave with its Poynting vector in the $z$ direction, and an electric field aligned with the $y$ axis varying as $E = E_0 cos(\omega t)$, the $y$ and $z$ components of the electron's momentum are respectively: $P_y = a_0m_ec\sin(\omega t)$ and $P_z = (a_0^2m_ec/4)\cos 2(\omega t)$. Hence for $a_0 >2$ or $I\lambda^2 >5.5\times10^{18}$W\textmu m$^2$cm$^{-2}$ the $z$ component of momentum exceeds that in $y$. For even higher intensities, the electron's final trajectory approaches that of the electromagnetic wave's Poynting vector, as $P_z \propto a_0^2/4$ while $P_y \propto a_0$. This electron acceleration mechanism is well known in high-intensity laser plasma physics, where the necessary intensities are readily achievable, and is commonly referred to as $\bf{j}\times \bf{B}$ acceleration.
 
\begin{figure}[!ht]
\centering
\includegraphics[width=.3\textwidth,trim=100mm 10mm 330mm 10mm,clip]{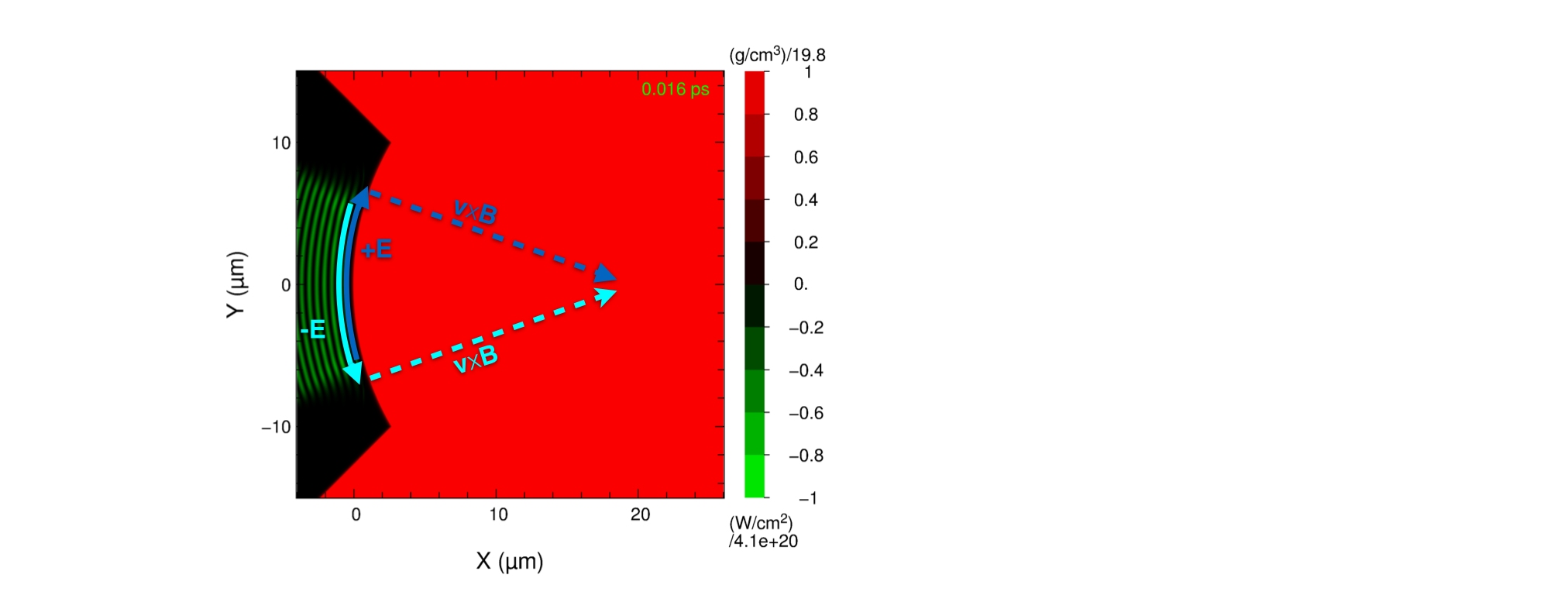}
\caption{The setup used in the PIC modelling and concept illustration. Green depicts the Poynting flux of the focussing incident laser which enters from the left, red is Au density. The solid blue upward arrow indicates the trajectory an electron might initially follow on one cycle of the laser when being principally accelerated by the laser's electric field. Wherever the electron is along its upward trajectory when it's speed reaches $\sim c$, the $\bf{v}\times \bf{B}$ component of the Lorentz force will be perpendicular to both the electric field and the target front surface (blue dashed line). In this example the electron reaches $c$ at the upward pointing solid blue arrowhead. As the radius of curvature of the laser and target front are 20 \textmu m, the electron is accelerated toward the focal point at y=0, x=20. Conversely on the next half-cycle of the laser the electric field accelerates the electron downwards (cyan solid arrow), but again the $\bf{v}\times \bf{B}$ component (cyan dashed arrow) accelerates the electron towards the focal point. It should be noted that were an electron to reach $c$ at \emph{any point} along its path over the target front surface, the $\bf{v}\times \bf{B}$ component would always be normal to its trajectory at that point.}
\label{concept}
\end{figure}

It follows from the above that for an electromagnetic wave which is non-planar, but smoothly varying in the $y$ direction (using the above coordinates), the electron's velocity vector will initially be locally parallel to the $\bf{E}$ field. Therefore wherever the electron is when its velocity reaches $\sim c$, and hence the $\bf{v}\times \bf{B}$ component of the Lorentz force becomes large, the electron will be accelerated in a direction $tan^{-1}(P_z/P_y) = tan^{-1}(a_0/4)$ with respect to the wave's \emph{local} Poynting vector, as shown in figure \ref{concept}. For an electromagnetic wave with a phase front of uniform curvature (a focussing, spherical wave), the Poynting vector at each location on the phase front points towards one point - the focal point of the wave/centre of curvature. Therefore if $a_0$ is sufficient, any electrons experiencing the focussing wave will be accelerated towards the wave's centre of curvature. This geometry occurs naturally with a focussing laser beam, prior to the beam reaching focus. Hence by combining a focussing laser beam with a solid target which has the same radius of curvature as that of the laser at the target front, but that is convex (as viewed by the laser), it should be possible, based on single electron dynamics, to create a \emph{convergent} or focussing fast-electron source. This interaction geometry is illustrated in figure \ref{concept}, with the Poynting flux shown in green while the target density is shown in red. Note the target front surface is placed well before the laser reaches peak focus, and the laser's electric field is parallel to the y-axis at y=0.

In order to investigate this interaction geometry, two dimensional simulations have been performed with the particle-in-cell code EPOCH \cite{0741-3335-53-1-015001}. They were initialised with 800 particles/species/cell, with convergence tests up to 1600. Using cubic spline particle shapes, numerical heating and grid convergence tests (minimum grid spacing  $1.6\times 10^{-3}$\textmu m) indicate that a spatial resolution of up to 150 times the Debye length (0.01\textmu m) is acceptable; over 150fs the plasma electrons numerically heated from an initial temperature of 89eV to 89.5eV (110eV) with collisions off (on) while the fast-electron generation and transport was quantitatively unaffected with respect to higher grid resolution runs. The timestep was 0.01 fs, while the simulated density was solid density Gold/Diamond-like-Carbon/DT with an assumed ionisation state of 10+/5+/1+. A density scale length of 0.1\textmu m was used at the target front surface, varying this altered the fast-electron temperature and absorption fraction. The laser wavelength was 1.064\textmu m and the peak time-averaged Poynting flux was $1\times 10^{20}$ Wcm$^{-2}$ at the target surface, although the scheme was robust over a range of relativistic intensities. Thermal boundary conditions were used. In order to prevent anomalous boundary effects associated with the fast-electron beam, the simulation was stopped before the beam reached the box edge.

\begin{figure*}[t!]
\centering
\includegraphics[width=1.\textwidth,trim=195mm 120mm 20mm 8mm,clip]{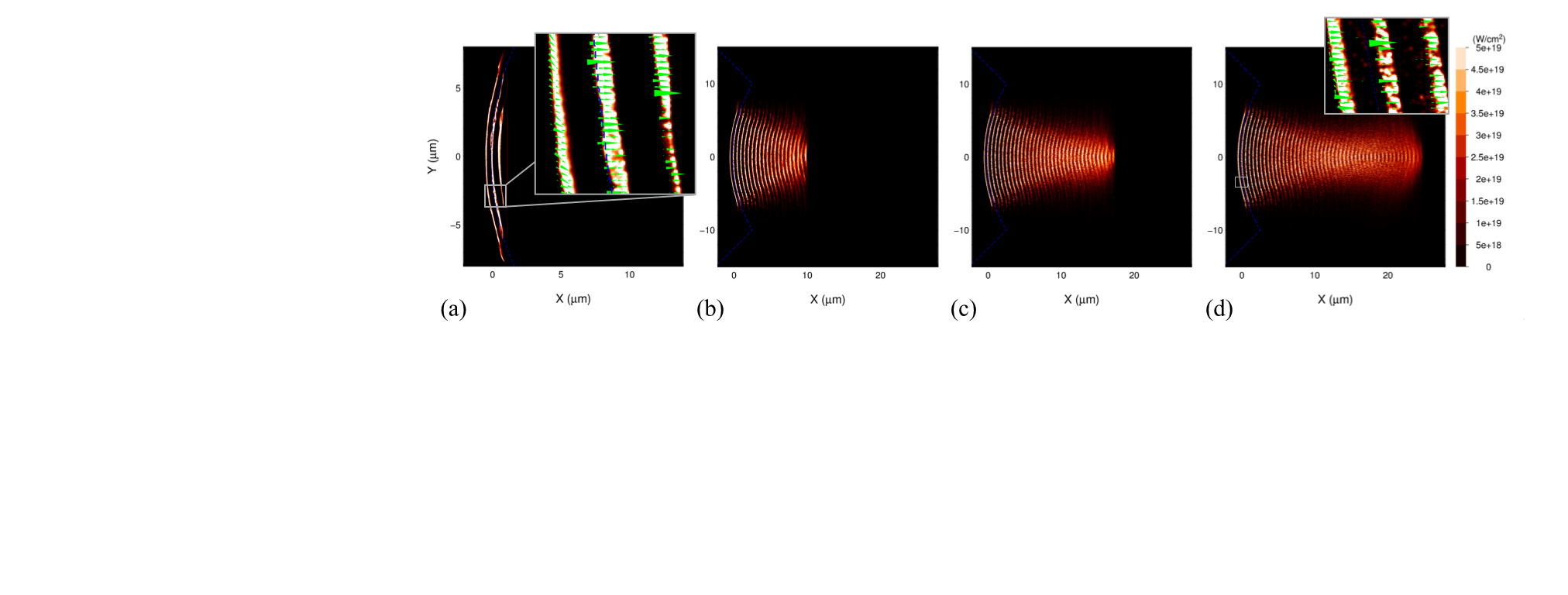}
\caption{(a) Fast-electron flux at 18 fs, note the spatial scale is smaller on this plot for clarity. the inset depicts the fast-electron flux vectors of a small region (grey box). (b) Fast-electron flux at 50 fs. (c) Fast-electron flux at 75 fs. (d) Fast-electron flux at 100 fs, inset depicts the fast-electron flux vectors of a small region (grey box). In all cases the blue dashed line denotes the initial position of the solid Au surface.}
\label{noIons}
\end{figure*}

In order to characterise the fast electron source with minimal additional complexity, initial EPOCH simulations were performed with all particle collisions off and no ion species. As EPOCH updates Maxwell's equations using the current, having no ions is equivalent to having infinitely massive ions. Figures \ref{noIons} (a)-(d) depict the flux of fast-electrons (defined throughout as those with kinetic energy greater that 0.5 MeV and less than 3MeV) induced by the laser plasma interaction depicted in figure \ref{concept} (solid Au target). The curved peaks in the flux intensity distribution have a spacing of 0.5 \textmu m, or half the laser wavelength, indicating absorption by the $\bf{j}\times \bf{B}$ mechanism. Figures \ref{noIons} (a)-(d) illustrate the propagation of a convergent fast electron beam within the solid density target. The radius of curvature of the target front and the laser are $20$\textmu m, while the fast electron beam reaches maximum focus at approximately this depth, as expected from the above single-electron dynamics arguments. The inset vector plots also confirm the directionality of the fast-electrons. Note that in the inset of fig. \ref{noIons} (a) the velocity vectors outside the solid surface (those to the left of the blue dashed line) are not locally perpendicular to the target surface. This is because at the time of this snapshot they have not yet received the full $\bf{j}\times \bf{B}$ push of this particular cycle of the laser, and hence are still predominately being accelerated down the target surface by the electric field.

The physics in the above PIC simulation is simplified by the effective use of infinitely massive ions and by omitting collisions. Nevertheless it confirms that in this limit, and for the conditions examined, the interaction is reasonably described by single electron dynamics which predicts both the temporal behaviour and trajectories of the fast electrons. Separate simulations (not shown) consistently confirmed that the focal length of the fast-electron beam approximately equals the radius of curvature of the laser/target front.

By neglecting ion motion, effects caused by deformation of the critical surface or expansion due to heating of the front surface ions are ignored. Deformation of the critical surface may de-optimise the interaction geometry, potentially degrading the fast-electron source directionality, reducing the focussed fast-electron beam intensity. Furthermore compression or expansion of the front surface is expected to change the fast-electron temperature.
  \begin{figure}[b!]
\centering
\subfigure[]{}\includegraphics[width=0.235\textwidth,trim=6mm 7mm 10mm 16mm,clip]{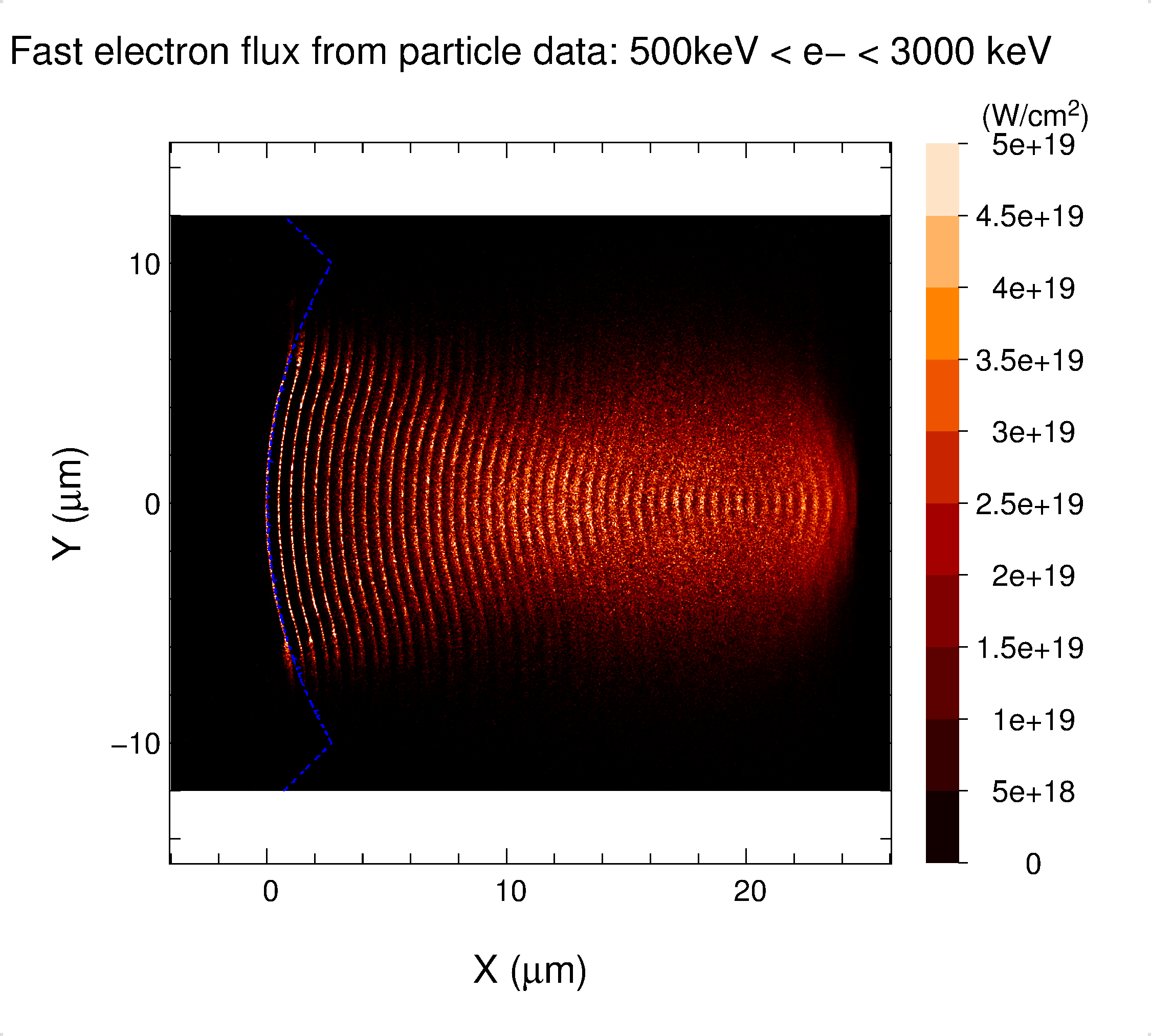}            
\subfigure[]{}\includegraphics[width=0.235\textwidth,trim=6mm 7mm 10mm 17mm,clip]{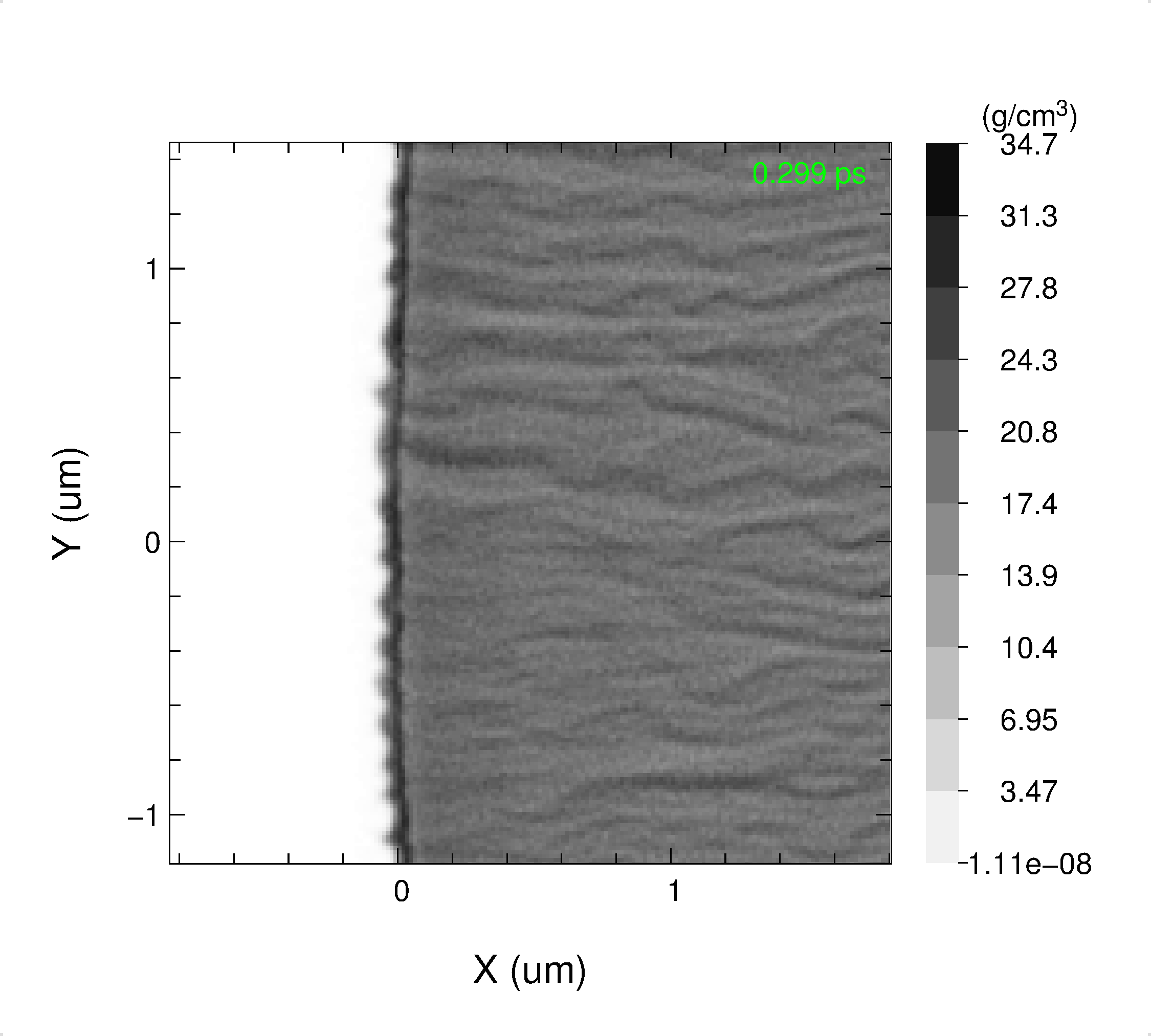}
\caption{(a) Fast-electron beam flux generated from simulated laser-solid interaction including ion motion but no collisions at 100 fs. (b) Zoomed in density plot of the target surface after 300fs.}
\label{ionMotion}
\end{figure}

Figure \ref{ionMotion} (a) depicts a simulation with is identical to that shown in fig. \ref{noIons} except that ion motion has now been enabled. It can be seen that for the timescales examined, the fast-electron source is similar to that generated when ion motion is prevented. Figure \ref{ionMotion} (b) depicts the target front surface at 300 fs, from an otherwise identical run but with reduced target depth for computational efficiency. The dense region of the target front surface is largely unperturbed, despite the effects of hole-boring and the fact that this interface is Rayleigh-Taylor \cite{Rayleigh:1900kx,Taylor:1950vn} unstable due to the opposing gradients in density and pressure at the target front surface. This indicates that little degradation in the source directionality due to deformation of the critical surface caused by ion motion would be expected up to this time. The principal difference observed when ion motion is included is a temporal variation in the fast-electron energy distribution function (spectrum). This is caused by the laser compressing the pre-imposed density ramp, steepening the density profile and reducing the absorption into fast-electrons. As the front surface ions heat up and expand, a shelf forms in the ion density profile \cite{Kemp:2008aa}. This causes an associated temporal variation in both the fraction of laser-light absorbed and the fast-electron energy distribution function. Despite the additional complexity caused by the ion motion, this result provides confirmation that a convergent fast-electron source is created using this novel interaction geometry. 

\begin{figure}[t!]
\centering
\includegraphics[width=0.495\textwidth,trim=29mm 3mm 170mm 18mm,clip]{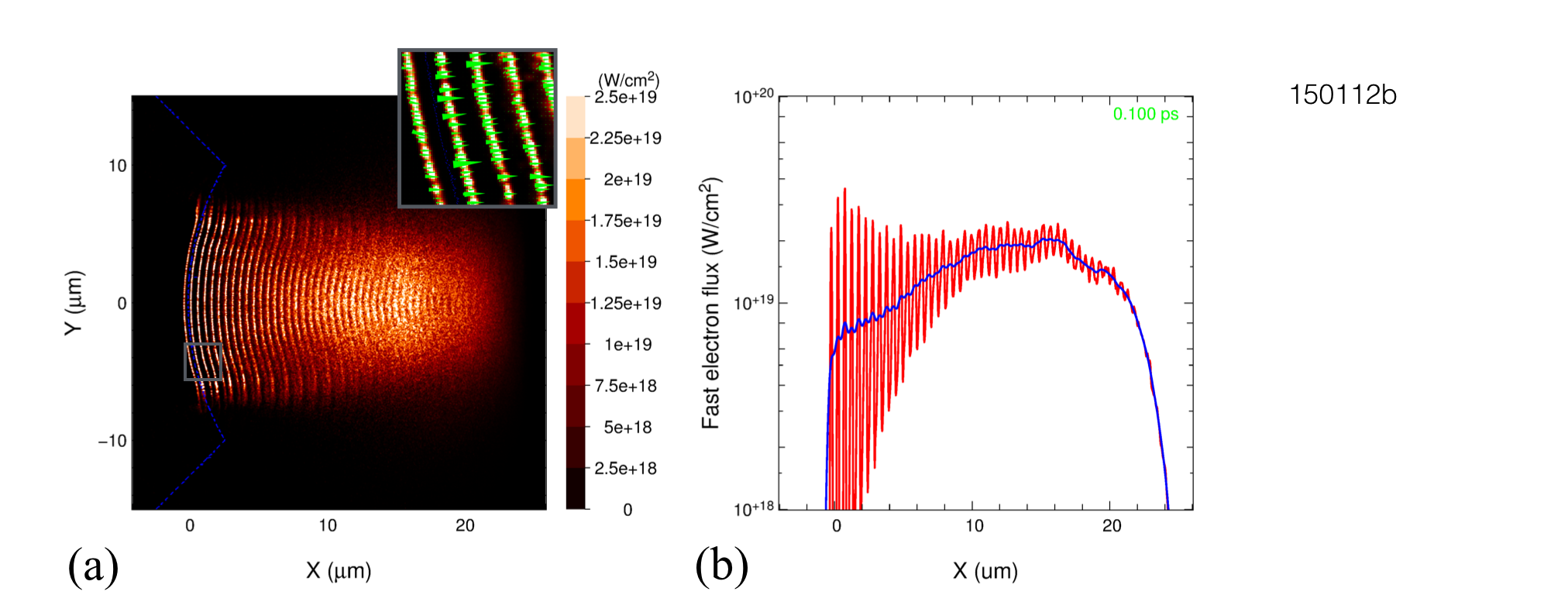}
\caption{(a) Fast-electron beam flux for PIC runs including both ion motion and collisions. Inset depicts fast-electron flux vectors within the region in the grey box shown in (a). (b) Red line is a lineout of fast-electron flux along the y=0 axis. Blue line is the same as the red, but smoothed.}
\label{collisions}
\end{figure}

In order to asses whether a convergent fast-electron beam propagates within the target from the convergent fast-electron source, the effects of collisions must be included during the transport of the fast-electrons through the target. Figure \ref{collisions} (a) depicts the $0.5-3$ MeV fast-electron flux from the same simulation as that shown in figs. \ref{noIons} (e) and \ref{ionMotion} (a) but particle collisions are now included. This has been simulated using a Coulomb logarithm of both 2 and 4, with little quantative difference observed. The fast-electron flux of the former is depicted in figure \ref{collisions}(a), again the inset shows the fast-electron flux vectors near the source point in approximately the correct direction for a focussing beam. In comparison to the previous runs it can be seen that the extent to which the beam converges is reduced, this is perhaps unsurprising as collisions will cause the beam to `bloom'. Figure \ref{collisions} (b) shows a lineout along the horizontal axis of fig. \ref{collisions}(a). The increase in fast-electron flux as a function of distance into the target indicates the beam is convergent,although this is convolved with temporal variations in the laser absorption and the fast-electron energy distribution function. In this collisional case a greater degree of velocity dispersion of the individual fast-electron bunches occurs, so that by 20 \textmu m the 2$\omega$ bunching of the individual fronts has largely disappeared. As the incident laser intensity is $1\times 10^{20}$Wcm$^{-2}$, and the fast-electron flux $\sim 1-3\times10^{19}$Wcm$^{-2}$ the fast-electron absorption fraction is of the order of 10-30\%. When the Au was transitioned to a fully ionised DT plasma at target depths greater than 1\textmu m, very similar results were obtained. If however the front surface of the target was made from either Diamond-like-Carbon or DT, deformation due to hole-boring caused the fast-electron beam to rapidly degrade.

Based on Bell et al's resistive self-collimation criterion \cite{Bell:2003lq}, it would be expected that the fast-electron beam generated using this scheme would be self-confined, however this is not observed in these simulations. Very large magnetic fields ($\sim 10^4$ T) are generated by the fast-electron beam, however these comprise many small-scale filamentary structures aligned with the local direction of the fast-electron beam propagation, the size of these are not inconsistent with the Weibel instability\cite{PhysRevLett.2.83}. The small scale magnetic field structures are at least in part due statistical noise; even with 1600 particles/species/cell full convergence of the magnetic field was not possible. The RMS amplitude of the magnetic field filaments is inversely proportional to particle number and is described by $B_{z_{RMS}} (T) = -171ln(pcls)+1611$, where $pcls$ is the total number of particles per cell. It is notable that hybrid-PIC simulations cause these filamentary field structures to largely disappear \cite{Cohen:2010:SLI:1767490.1767659}, furthermore when smoothed in post processing, the filaments merged into a large-scale azimuthal field with the correct geometry to self-collimate. It should be noted that despite the above, quantitative convergence of the fast-electron flux was demonstrated with $\sim$ 400 electrons and 50ions per cell. 

A convergent fast-electron beam has the potential to improve the energetics of the fast-ignitior scheme. By reducing the fast-electron beam radius, the hotspot mass is reduced and hence the required ignitor beam energy reduced. Furthermore by focussing the fast-electron beam it may be possible to use a lower laser intensity in order to heat the hotspot within the inertial confinement time, this would reduce the fast-electron temperature, improving collisional coupling. Whether the scheme outlined in this letter will translate into a viable ignitor beam is as yet unknown. This is due to the greatly increased temporal and spatial scales required to impart sufficient energy into the hotspot and simulate the anticipated offset from the cone-tip to the compressed DT fuel. In principal this fast-electron focussing scheme should be equally applicable to three spatial dimensions, here optimisation might be achieved by employing a radially polarised laser beam focussing towards a convex target. These issues will be addressed in future work.

In conclusion, a novel scheme for the generation of a convergent fast-electron source has been outlined. Proof of principal two dimensional PIC modelling has shown that this scheme works for the spatial and temporal scales examined. Furthermore it is shown that the fast-electron beam can propagate and focus within a solid density plasma.

The author would like to thank: A.R. Bell, A.P.L. Robinson, H. Schmitz and M. Sherlock for useful discussions, R.M.G.M Trines and P.A. Norreys for their support, the EPOCH collaboration (funded by UK EPSRC grants EP/G054950/1, EP/G056803/1 and EP/G055165/1), and the SCARF computing clusters. Access to HECToR and ARCHER computing clusters was funded by the Plasma HEC Consortium on EPSRC grant number EP/L000237/1. The author was funded by the Science and Technology Facilities Council, UK.  

%

\end{document}